\documentstyle[sprocl]{article}

\input{psfig}

\begin{document}

\title{SUBSTRUCTURE IN CDM HALOS AND THE HEATING OF STELLAR DISKS}

\author{Julio F. Navarro}

\address{Department of Physics and Astronomy, \\University of Victoria, \\Victoria,
BC, V8P 1A1, Canada}

\maketitle\abstracts{
Numerical simulations have revealed the presence of long-lived substructure in
Cold Dark Matter (CDM) halos. These surviving cores of past merger and accretion
events vastly outnumber the known satellites of the Milky Way.  This finding has
prompted suggestions that substructure in cold dark matter (CDM) halos may be
incompatible with observation and in conflict with the presence of thin,
dynamically fragile stellar disks. N-body simulations of a disk/bulge/halo model
of the Milky Way that includes several hundred dark matter satellites with
masses, densities and orbits derived from high-resolution cosmological CDM
simulations indicate that substructure plays only a minor dynamical role in the
heating of the disk. This is because the orbits of satellites seldom take them
near the disk, where their tidal effects are greatest. We conclude that
substructure might not preclude virialized CDM halos from being acceptable hosts
of thin stellar disks like that of the Milky Way.}

\section{Introduction}
One recent highlight of cosmological N-body simulations has been the discovery
that during the merger and accretion events that characterize the assembly of
dark matter halos the central regions of accreted halos may survive for several
orbital times as dynamically distinct, self-bound entities in the parent halo
(Klypin et al 1999, Moore et al 1999, hereafter K99 and M99, respectively).  The
population of surviving halo cores, or ``subhalos'', typically contributes less
than $\sim 10\%$ of the total mass of the system, with the bulk of the mass in a
smooth monolithic structure, as envisioned in the pioneering analytical work of
White \& Rees (1978). Despite the small fraction of the total mass they make up,
at any given time a large number of subhalos are expected within the virialized
region of a cold dark matter halo. For example, M99 find that up to $500$
satellites with circular velocities exceeding $\sim 10$ km/s may have survived
within $\sim 300$ kpc from the center of the Galaxy.  Comparing this with the
dozen or so known Milky Way satellites of comparable velocity dispersion implies
that most subhalos must have failed to form a significant number of stars.

Even if luminous galaxies fail to ``turn on'' in most subhalos a potential
difficulty has been cited by M99: the fluctuating gravitational potential
induced by the clumpy structure of the halo may act to heat and thicken fragile
stellar disks beyond observational constraints.
I present here the results of an attempt to quantify the effects of substructure
on the dynamical evolution of thin stellar disks embedded in dark matter
halos. 

\section{Substructure in CDM halos}

The mass function of substructure halos, their internal structure, and the
parameters of their orbits are the main properties of the subhalo population
that determine the tidal effects of substructure on stellar disks.  The
short-dashed curves in Figure 1 (left panel) show the subhalo velocity function
corresponding to galaxy-sized dark matter halos formed in the $\Omega=1$ CDM and
in the $\Lambda$CDM cosmogonies.  Circular velocity (instead of mass) is used to
characterize subhalos because of its weaker dependence on the exact way in which
substructure is identified. Still, circular velocities do change as a function
of radius from the center of a subhalo, and there is no unique way of defining
subhalo circular velocities. Figure 1 reports results for $V_{\rm peak}$ and
$V_{\rm outer}$, which correspond to the maximum circular speed within the
subhalo, and to its value at the outermost bound radius, respectively. Scaled to
the virial velocity of the halo, the substructure velocity function is roughly
independent of the mass of the parent halo (M99) and of the cosmological
parameters (K99).  The results of K99 are shown by the dot-dashed line in Figure
1 (left) and are in reasonable agreement with our determination.

\begin{figure}[t]
%\rule{5cm}{0.2mm}\hfill\rule{5cm}{0.2mm}
%\vskip 2.5cm
%\rule{5cm}{0.2mm}\hfill\rule{5cm}{0.2mm}
\psfig{figure=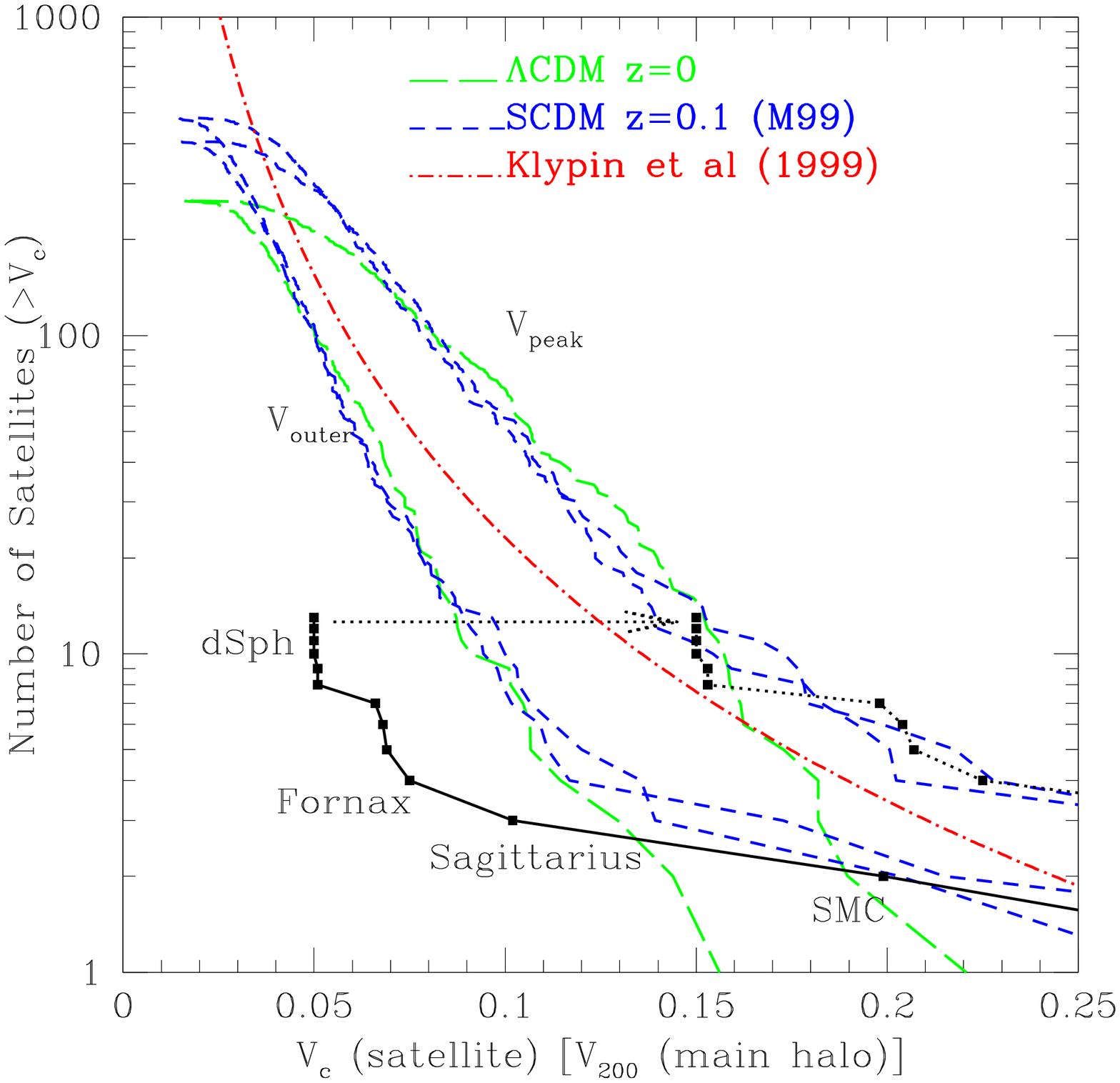,height=2.25in}
\vskip -5.75cm
\hskip 6.cm
\psfig{figure=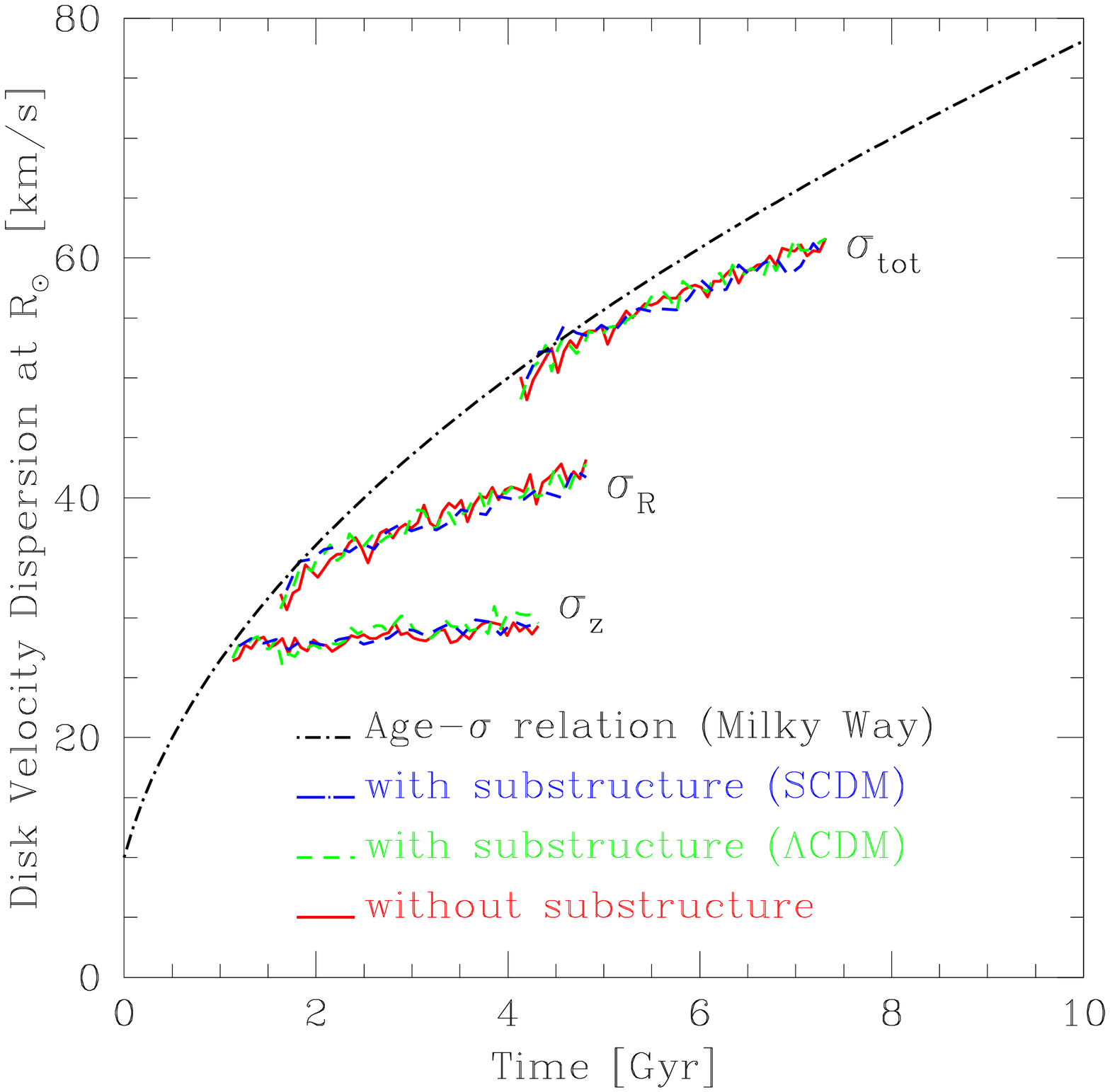,height=2.25in}
\caption{Left:
Cumulative circular velocity function of substructure halos.  Velocities are
normalized to the circular velocity of the parent halo measured at the virial
radius. Only subhalos within the virial radius of the main halo are included.
Right: Stellar disk velocity dispersion as a function of time in our Milky Way
model compared with the age-velocity dispersion relation in the solar
neighborhood, as compiled recently by Fuchs et al (2000, dot-dashed curve).
\label{fig:nvcirc}}
\end{figure}

How does the subhalo velocity function compare with the observed number of
satellites in the vicinity of the Milky Way? The solid squares in Figure 1
(joined by a solid line) illustrate the cumulative number of known Milky Way
satellites as a function of the circular velocity of their halos, as plotted by
M99. Here, circular velocities are derived for the halos of dwarf spheroidals
assuming that stars in these systems are on isotropic orbits in isothermal
potentials. This is a plausible, but nevertheless questionable, assumption. Dark
halos differ significantly from simple isothermal potentials, and numerical
simulations indicate that circular velocities decrease substantially near the
center. If stars populate the innermost regions of subhalos their velocity
dispersions may substantially underestimate the subhalo peak circular
velocities. This has been discussed by White (2000), who finds, using the mass
model proposed by Navarro, Frenk \& White (1997, NFW), that dwarf spheroidals
may plausibly inhabit potential wells with circular velocities up to a factor of
$3$ times larger than inferred under the isothermal assumption.  Such correction
may reconcile, at the high mass end, the Milky Way satellite velocity function
with the subhalo function, as shown by the dotted line in Figure 1. Thus, the
possibility remains that the number of {\it massive} satellites expected in the
CDM scenario might not be in gross conflict with observation.

On the other hand, this does not erase the large number of low mass ``dark''
satellites that should inhabit the main halo. Can thin stellar disks survive
unscathed in this clumpy environment? The tidal heating rate by substructure can
be shown to scale as $dE/dt \propto \int n(m_{s}) \, m_{s}^2 \, dm_{s}$, where
$m_s$ is the subhalo mass (White 2000). Since, according to numerical
simulations, $n(m_s)\propto m_s^{-1.8}$ (Ghigna et al 2000, Springel et al
2000), tidal heating by substructure is dominated by the few most massive
subhalos. One consequence of this result is that stochastic effects are expected
to dominate the heating rate and a statistically significant sample of systems
should be simulated in order to obtain conclusive results. The results reported
here (which are based on a couple of realizations) should then be viewed as
preliminary until confirmed by further studies.

\section{The Evolution of Stellar Disks in a $\Lambda$CDM halo}

In order to assess the dynamical influence of subhalos on stellar disks we have
carried out a series of numerical simulations that follow the evolution of a
disk of particles within a dark matter halo with substructure similar to that of
the halos shown in Figure 1 (left).  We choose parameters so that the model
reproduces many of the observational characteristics of the Milky Way galaxy, as
in Vel{\'{a}}zquez \& White (1999). The galaxy model is first evolved without
substructure halos in an attempt to assess the deviations from equilibrium
induced by noise in the particle distribution. N-body disks are notoriously
unstable, and substantial numbers of particles are needed to ensure stability
over several rotation periods. The simulations we report here use $40,000$
particles in the disk, $13,333$ in the bulge, and $2.2 \times 10^6$ in the halo,
all of equal mass, $m_p=1.4 \times 10^{6} M_{\odot}$.  Within this system we
insert substructure halos, using a procedure which ensures that the subhalos
have approximately the same masses, densities and orbits as those in the CDM
halos.

We evolve the disk model first for about $3.5$ Gyrs without including
substructure in order to quantify the heating rate due to noise in the particle
distribution. The ($R$, $z$, $\phi$) velocity dispersions of disk particles at
the solar circle (i.e., $8.5$ kpc from the center) grow from ($31$, $27$, $26$)
km s$^{-1}$ to ($43$, $30$, $33$) km s$^{-1}$ over the same period (see solid
lines in Figure 4). Quantifying this heating rate by the usual expression
$\sigma_{tot}^2=\sigma_0^2+Dt$, we find $D\sim 200$ km$^2$ s$^{-2}$ Gyr$^{-1}$,
about a factor of two less than inferred for stars in the solar neighborhood
from the age-velocity dispersion (Fuchs et al 2000).

The two-body heating rate in our equilibrium stellar disk model is thus low
compared with the actual heating experienced by stars in the disk of the
Galaxy. This implies that our disk model is stable enough to verify numerically
whether substructure in the halo leads to heating rates inconsistent with
observational constraints. This is shown by the dashed lines in Figure 4, which
show the evolution of the disk velocity dispersion when the substructure halos
are added to the system. Clearly, the heating rate is approximately the same
with and without substructure, a result that may be traced to the fact that
there are no satellites more massive than $10^{11} \, M_{\odot}$ and that their
orbits seldom take them near the disk. If substructure in the halos we consider
is representative of galaxy-sized CDM halos (and we have no reason to suspect it
is not), this would imply that tidal heating rates of thin stellar disks by
substructure halos may be consistent with the observational evidence.

\section{Discussion}
These results suggest that concerns regarding excessive tidal heating of thin
stellar disks by substructure in CDM halos may be less serious than previously
thought.  We conclude that the substructure observed in virialized CDM halos is
not clearly inconsistent with the existence of thin stellar disks such as that
of the Milky Way. These conclusions are subject to a number of caveats. The most
obvious one is that our study explores only two numerical realizations of a disk
galaxy within clumpy dark matter halos, and it is always hazardous to
extrapolate from such a small number of cases. Our study does show, however,
that it is possible at least in some cases to maintain a stellar disk in spite
of substructure. A second caveat is that we have explored a model motivated by
the present-day structure of the Milky Way and by the $z=0$ substructure of a
CDM halo. Models that take into account the ongoing formation of the disk and a
more realistic treatment of the evolution of substructure are clearly desirable
in order to refine the conclusions presented here.

Finally, as discussed by Navarro \& Steinmetz (2000), it is quite difficult to
account simultaneously for the masses, luminosities, rotation speeds, and
angular momenta of galaxy disks in cosmogonies such as CDM, where much of the
mass of a virialized halo tends to be assembled through a sequence of
mergers. Until these issues are fully resolved it would be premature to extend a
clean bill of health to the CDM paradigm regarding the formation and evolution
of spiral galaxies like our own Milky Way.

\section*{Acknowledgments}
I thank my collaborators, Andreea Font, Joachim Stadel and Tom Quinn for
allowing me to present the results of our collaboration. This research has been
supported by CIAR, CFI, NSERC and the Alfred P. Sloan Foundation.

\end{document}